# Fast-forward scheme reexamined: Choice of time and quantization

Sumiyoshi Abe [1-4]

[1] *Department of Physics, College of Information Science and Engineering, Huaqiao University, Xiamen 361021, China*
[2] *Institute of Physics, Kazan Federal University, Kazan 420008, Russia*
[3] *Department of Natural and Mathematical Sciences, Turin Polytechnic University in Tashkent, Tashkent 100095, Uzbekistan*
[4] *ESIEA, 9 Rue Vesale, Paris 75005, France*

ABSTRACT

The fast-forward scheme for accelerating time evolution of quantum states through change of time is reexamined. Dirac's homogeneous formalism for classical dynamics and canonical quantization warrant that the Schrödinger equation is covariant under reparametrization of time. From this, it is concluded that the scheme does not attain its objective.

*Keywords:*
fast-forward problem
reparametrization of time
homogeneous formalism
canonical quantization

Accelerating quantum state evolution is of fundamental interest from the viewpoints of adiabatic computing, cooling of atoms, quantum transport and quantum thermodynamics, to name a few. There are at least two approaches: one is "shortcuts to adiabaticity" [1-3] and the other is "fast-forward scheme" [4,5]. Both of them have recently been attracting much attention. Also, a discussion has been made in Ref. [6] about a possible relationship between the two.

What we reexamine here is concerned with the fast-forward scheme. For this purpose, we start our discussion with classical dynamics, in order to clarify how choice of time affects canonical quantization. In particular, we base our starting point in Dirac's homogeneous formalism for classical dynamics. We see that the Schrödinger equation is covariant under reparametrization of time, and accordingly the scheme does not accelerate time evolution of quantum states.

Consider the Schrödinger equation: $i\hbar(\partial/\partial t)|\psi(t)\rangle = \hat{H}(t)|\psi(t)\rangle$. Here, the system Hamiltonian may have explicit time dependence that is not necessarily slow, in general. In the fast-forward scheme [4,5], reparametrization of time is performed for the equation. A simple case is the rescaling $t \to \alpha t$, where $\alpha$ is a positive dimensionless constant. The new clock goes faster than the original one if $\alpha > 1$. The scheme rewrites the Schrödinger equation as follows: $i\hbar(\partial/\partial t)|\psi_\alpha(t)\rangle = \hat{H}_\alpha(t)|\psi_\alpha(t)\rangle$, where $|\psi_\alpha(t)\rangle = |\psi(\alpha t)\rangle$ and $\hat{H}_\alpha(t) = \alpha\hat{H}(\alpha t)$. Then, the fast-forwarded state $|\psi_{FF}(t)\rangle$ satisfying $i\hbar(\partial/\partial t)|\psi_{FF}(t)\rangle = \hat{H}_{FF}(t)|\psi_{FF}(t)\rangle$ is constructed in terms of $|\psi_\alpha(t)\rangle$ in the unitarily equivalent manner, where $\hat{H}_{FF}(t)$ may be refereed to as the "fast-forwarding Hamiltonian". In the analysis [4,5], the position representation plays an outstanding role. Thus, the scheme regards the Schrödinger equation in the position representation with a usual partial differential equation, reparametrizes time and then changes the equation to that in original conventional time through the redefinition of the wavefunction as well as controls of the phase factor and the Schrödinger operator. The resulting wave equation is not linear any more.

Now, the following question is asked: is it allowed to change time in the Schrödinger equation as in a usual partial differential equation in the above-mentioned way? This question is deeply related to the concept of canonical quantization that has to be performed after choice of time. The purpose of this note is to reexamine such an issue for the fast-forward scheme. At the same time, this seems to give an opportunity to recall Dirac's homogeneous formalism for nonrelativistic classical dynamics [7-9], although the formalism becomes clearer in relativistic dynamics where the space and time coordinates are treated on an equal footing and the manifest reparametrization invariance is realized.

The original idea of the homogeneous formalism is to make conventional time, $t$, as a dynamical variable

$$t = T(\tau), \tag{1}$$

where $\tau$ is an arbitrary parameter satisfying a natural condition

$$\frac{dT(\tau)}{d\tau} > 0. \tag{2}$$

However, the formalism can also be applied to the problem of choosing time in canonical quantization. Let us consider the action in conventional time $t$:

$$I = \int dt\, L(t, x, \dot{x}), \tag{3}$$

where the overdot stands for the derivative with respect to $t$. The Euler-Lagrange equation is then given by

$$\frac{d}{dt}\left(\frac{\partial L}{\partial \dot{x}}\right) - \frac{\partial L}{\partial x} = 0. \tag{4}$$

The canonical momentum conjugate to $x$ is $p = \partial L / \partial \dot{x}$, and the Hamiltonian is given by the Legendre transformation,

$$H(t, x, p) = \dot{x}\, p - L, \tag{5}$$

which generates $t$-evolution of the system.

On the other hand, let us write the action in terms of $\tau$ as follows:

$$I = \int d\tau\, \tilde{L}(T, \xi, T', \xi'), \tag{6}$$

where $\xi(\tau) \equiv x(T(\tau))$ and the prime denotes the derivative with respect to $\tau$. Therefore, comparing Eq. (3) in its rewritten form

$$I = \int d\tau\, T'\, L(T, \xi, \xi'/T') \tag{7}$$

with Eq. (6), we have the following relation between the two Lagrangians:

$$\tilde{L}(T,\xi,T',\xi') = T' L(T,\xi,\xi'/T').  \qquad (8)$$

Clearly, Eq. (3) is reproduced in the gauge $T = \tau (= t)$. Equation (8) implies that the new Lagrangian, $\tilde{L}$, is a homogeneous function of degree one. Therefore, from Euler's theorem, it follows that

$$T'\frac{\partial \tilde{L}}{\partial T'} + \xi'\frac{\partial \tilde{L}}{\partial \xi'} = \tilde{L}.  \qquad (9)$$

Taking the derivative of this equation with respect to $\tau$, we have

$$T''\frac{\partial \tilde{L}}{\partial T'} + T'\frac{d}{d\tau}\left(\frac{\partial \tilde{L}}{\partial T'}\right) + \xi''\frac{\partial \tilde{L}}{\partial \xi'} + \xi'\frac{d}{d\tau}\left(\frac{\partial \tilde{L}}{\partial \xi'}\right)$$
$$= T'\frac{\partial \tilde{L}}{\partial T} + \xi'\frac{\partial \tilde{L}}{\partial \xi} + T''\frac{\partial \tilde{L}}{\partial T'} + \xi''\frac{\partial \tilde{L}}{\partial \xi'},  \qquad (10)$$

which gives rise to

$$\frac{d}{d\tau}\left(\frac{\partial \tilde{L}}{\partial T'}\right) - \frac{\partial \tilde{L}}{\partial T} = 0,  \qquad (11)$$

where the Euler-Lagrange equation for $\xi$,

$$\frac{d}{d\tau}\left(\frac{\partial \tilde{L}}{\partial \xi'}\right) - \frac{\partial \tilde{L}}{\partial \xi} = 0,  \qquad (12)$$

derived from the action in Eq. (6) has been used. Equation (11) implies that conventional time in Eq. (1) can be treated as another dynamical variable parametrized by $\tau$. In other words, the formalism shows how evolution of a dynamical system is described by various definitions of time. Time as a dynamical degree of freedom is however redundant. To see it, let us consider the canonical momenta conjugate to $\xi$ and $T$,

$$\pi = \frac{\partial \tilde{L}}{\partial \xi'}, \qquad \pi_T = \frac{\partial \tilde{L}}{\partial T'}, \tag{13}$$

respectively. Then, from Eq. (9), it follows that the constraint

$$T'\pi_T + \tilde{H} = 0, \tag{14}$$

identically holds, where

$$\tilde{H}(T, \xi, \pi) = \xi'\pi - \tilde{L}$$

$$= T'H(T, \xi, \pi) \tag{15}$$

is the Hamiltonian generating $\tau$-evolution of the system. It should be noted that the functional form of $H(T, \xi, \pi)$ is the same as that of $H(t, x, p)$ in Eq. (5), as will explicitly be seen in the discussion in the second-last paragraph.

In the particular gauge $T = \tau (= t)$, the constraint in Eq. (14) becomes $\pi_t + H(t, x, p) = 0$. Upon canonical quantization [see Eq. (16) below], such a constraint is regarded as the supplementary condition on the physical state, say $|\Psi\rangle$ [10]:

$\left(\hat{\pi}_t + \hat{H}(t)\right)|\Psi\rangle = 0$, where $\hat{H}(t) \equiv H(t, \hat{x}, \hat{p})$ is the Hamiltonian operator. Here, $t$ is not treated as an observable and accordingly the quantization condition on $\pi_t$ and $t$ does not exist [11] although the problems of time operators have repeatedly been discussed in the literature. The "position-time representation" of the supplementary condition, $\langle x, t|\left(\hat{\pi}_t + \hat{H}(t)\right)|\Psi\rangle = 0$, formally leads to the "Schrödinger-like" wave equation: $i\hbar \partial \Psi(x,t)/\partial t = H(t, x, -i\hbar \partial/\partial x) \Psi(x,t)$, where $\langle x, t|\hat{p} = -i\hbar \partial/\partial x \langle x, t|$, $\langle x, t|\hat{\pi}_t = -i\hbar \partial/\partial t \langle x, t|$, and $\Psi(x,t) = \langle x, t|\Psi\rangle$. Here, again we emphasize that time is not regarded as an observable, and therefore, $|x, t\rangle$ is the eigenstate of the position operator, $\hat{x}$, but not of any time operator. The wave equation of this familiar form is obtained only in the particular gauge, $T = \tau (=t)$, and such a gauge fixing is performed before quantization.

Now, the quantization condition is

$$\left[\hat{p}, \hat{x}\right] = -i\hbar, \tag{16}$$

$$\left[\hat{\pi}, \hat{\xi}\right] = -i\hbar, \tag{17}$$

and, correspondingly, the Schrödinger equation takes the forms

$$i\hbar \frac{\partial}{\partial t}|\psi(t)\rangle = \hat{H}(t, \hat{x}, \hat{p})|\psi(t)\rangle, \tag{18}$$

$$i\hbar\frac{\partial}{\partial\tau}|\phi(\tau)\rangle = T'\hat{H}(T,\hat{\xi},\hat{\pi})|\phi(\tau)\rangle, \tag{19}$$

in *t* and $\tau$, respectively.

The crucial point in Eq. (19) is the presence of the prefactor $T'$ on the right-hand side, *which is missing in the fast-forward scheme*. This factor assures the covariance of the Schrödinger equation under reparametrization of time. Therefore, up to irrelevant phases, $|\phi(\tau)\rangle$ and $|\psi(T(\tau))\rangle$ are equivalent.

As an explicit example illustrating the above discussion, let us consider a typical system whose action is given by $I = \int dt\{(1/2)m\dot{x}^2 - V(t,x)\}$, where *m* and *V* are the mass of the particle and a time-dependent potential, respectively. The canonical momentum conjugate to *x* is $p = m\dot{x}$, and the Hamiltonian is given by $H(t,x,p) = p^2/(2m) + V(t,x)$ that generates *t*-evolution of the system. In this example, Eq. (8) becomes $\tilde{L} = m\xi'^2/(2T') - T'V(T,\xi)$, which clearly satisfies Euler's theorem in Eq. (9). Let us eliminate the redundant degree of freedom and regard Eq. (1) as a change of the time parameter, that is, $T(\tau)$ is not a dynamical variable. The canonical momentum conjugate to $\xi$ is $\pi = m\xi'/T'$, and the Hamiltonian is given by $\tilde{H}(T,\xi,\pi) = T'H(T,\xi,\pi)$ with $H(T,\xi,\pi) = \pi^2/(2m) + V(T,\xi)$ having the same functional form as $H(t,x,p)$. Under the quantization condition in Eq. (17), the Schrödinger equation in $\tau$ reads $i\hbar(\partial/\partial\tau)|\phi(\tau)\rangle = T'\{\hat{\pi}^2/(2m) + V(T,\hat{\xi})\}|\phi(\tau)\rangle$. In the position representation with the basis $\langle x|$, both $\hat{x}$ and $\hat{\xi}$ become *x*, and $\hat{p}$ and $\hat{\pi}$ the same differential operator $-i\hbar\partial/\partial x$, respectively, since these observables do not depend on time in the Schrödinger picture. In the simplest case of $t = T(\tau) = \tau/\alpha$ mentioned in the very beginning of the present note, the factor $1/\alpha$ cancels out in the Schrödinger equation, exhibiting its covariance under the reparametrization of time.

In conclusion, the fast-forward scheme in quantum mechanics has been reexamined by use of the homogeneous formalism for classical dynamics and canonical quantization. Since the Schrödinger equation is covariant under reparametrization of time, the scheme does not attain its objective.

**Acknowledgments**

This work has been supported in part by a grant from National Natural Science Foundation of China (No. 11775084), the Program of Fujian Province, and the Program of Competitive Growth of Kazan Federal University from the Ministry of Education and Science of the Russian Federation, to all of which the author gratefully acknowledges.